\documentclass[twocolumn,showpacs,preprintnumbers,amsmath,amssymb,10pt]{revtex4}

\usepackage{psfig}
\usepackage{graphicx}
\usepackage{dcolumn}
\usepackage{bm}

\newcommand{\s}{\ensuremath{\psi(t,r)}}

\newcommand{\T}{\ensuremath{\theta}}

\newcommand{\e}{equation} 
\newcommand{\M}{\ensuremath{{\cal M}}}

\newcommand{\X}{\ensuremath{{\cal X}}}

\newcommand{\dw}{\ensuremath{{d\Omega^2_{N-2}}}}

\newcommand{\rdot}{\ensuremath{\dot{R}}}
\newcommand{\rddot}{\ensuremath{\ddot{R}}}

\newcommand{\F}{\ensuremath{F(r)}}
\newcommand{\f}{\ensuremath{f(r)}}

\begin{document}
\preprint{}
\title{Spherical Dust Collapse in Higher Dimensions}
\author{Rituparno Goswami}
\email{goswami@tifr.res.in}
\author{Pankaj S Joshi}
\email{psj@tifr.res.in}
\affiliation{ Department of Astronomy and Astrophysics\\ Tata 
Institute of Fundamental Research\\ Homi Bhabha Road,
Mumbai 400 005, India}

\begin{abstract}
We consider here the question if it is possible to recover cosmic 
censorship when a transition is made to higher dimensional spacetimes,
by studying the spherically symmetric dust collapse in an arbitrary
higher spacetime dimension. It is 
pointed out that if only black holes are to result as end state of a 
continual gravitational collapse, several conditions must be imposed on 
the collapsing configuration, some of which may appear to be restrictive, 
and we need to study carefully if these can be suitably 
motivated physically in a realistic collapse scenario. It would appear 
that in a generic higher dimensional dust collapse, both black holes 
and naked singularities would develop as end states as indicated by 
the results here. The mathematical approach developed here generalizes
and unifies the earlier available results on higher dimensional dust 
collapse as we point out. Further, the dependence of black hole or naked 
singularity end states as collapse outcomes, on the nature of the 
initial data from which the collapse develops, is brought out explicitly 
and in a transparent manner as we show here. Our method 
also allows us to consider here in some detail the genericity and 
stability aspects related to the occurrence of naked singularities in 
gravitational collapse.

\end{abstract}
\pacs{04.20.Dw, 04.20.Cv, 04.70.Bw}
\maketitle

\section{Introduction} 
A considerable debate has continued in recent years
on the validity or otherwise of the cosmic censorship conjecture (CCC) in
black hole physics 
(see e.g. [1-7] 
for some recent reviews). The reason
for this interest is that CCC is fundamental to many aspects of theory and
applications of black hole physics, and the astrophysical implications 
resulting
from a continual gravitational collapse of a massive star which has
exhausted its nuclear fuel. As of today, no theoretical proof or any
satisfactory mathematical formulation of CCC is available, where as many
collapse scenarios have been analyzed and worked out wherein the collapse
end state is either a {\it black hole} (BH) or a {\it naked singularity}
(NS), depending on the nature of the initial data from which the collapse
evolves developing from a regular initial state to the final super-dense
state. This has important astrophysical significance for the reason that
naked singularities may have observational properties which could 
be radically different from those of a black hole.

A note-worthy suggestion that has emerged towards a possible theoretical
formulation of CCC is that any naked singularities resulting from matter
models which may also develop singularities in special relativity, should
not be regarded as physical 
(see e.g. [3]). 
Clearly, it will   
require a serious effort to cast such a possibility into a mathematical 
statement and a
proof for CCC. Also, it may not be easy to discard completely all the
matter fields such as dust, perfect fluids, and matter with various other 
reasonable equations of state, which have been studied and used 
extensively in relativistic astrophysics for a long time.

Another possibility that indeed appears worth exploring
is we may actually be living in a higher dimensional spacetime
universe. The
recent developments in string theory and other field theories strongly   
indicate that gravity is possibly a higher dimensional interaction, which   
reduces to the general relativistic description at lower energies. Hence,
while CCC may fail in the four-dimensional manifold of general 
relativity, it
may well be restored due to the extra physical effects arising from our   
transition itself to a higher-dimensional spacetime continuum. Such 
considerations would inspire a study of gravitational collapse 
in higher dimensions. 
From such a perspective, many papers have 
reported results in recent years on spherically symmetric collapse of 
dust (where the pressures are identically taken to be zero) in higher 
dimensions
[8-11]. 
The recent revival of interest in this problem is
motivated to an extent by the Randall-Sundrum brane-world 
scenario 
[12]. 
Various authors 
considered specialized and different subcases
of the general problem of dust collapse in higher spacetime
dimensions. For example, the marginally bound case
in a general spacetime dimension was studied in [9], 
whereas the same case was studied in [10], 
however, with an additional and physically motivated restriction 
on the initial density profiles, namely that the first derivative
of the initial density distribution for the collapsing cloud must
be vanishing. Also, the non-marginally bound case, however, 
with the geometric assumption that spacetime is self-similar,  
was examined in [11] for a five-dimensional model.

Such studies do provide us with an idea of what is possible in 
gravitational
collapse as far as its end state spectrum is concerned.
It is obvious from the work in this area so far that any 
possible proof of CCC must
be inspired by such additional physical inputs (one of these being a 
possible transition to higher dimension) into our current framework
of thinking. Any such alternatives would be worth exploring due to the
fundamental significance of CCC in black hole physics. 
The point is, if naked singularities
did indeed develop in realistic gravitational collapse of massive
objects, they may have properties which would be rather different 
from those of black holes both theoretically as well as observationally, 
and a comparison of these two cases may
prove to be quite interesting.

From such a perspective, we investigate here the issue of BH/NS 
end states in a higher dimensional dust collapse in some detail, in order 
to bring out in a transparent manner the effect of dimensions on the 
final fate of the evolution of the matter cloud which collapses from 
a given regular initial data. A spherically symmetric collapse is 
considered in $N\ge4$ dimensions
and the matter form is chosen to be dust as this is a well-studied case. 
In fact, there have been suggestions in
the past that at the vicinity of the collapse, the in-falling velocity 
of the matter shells could be so high that the effects of pressures are 
negligible and hence dust may be a good approximation 
\cite{haga}. Apart from that, dust collapse is worth investigation
as it has continued to serve as a basic paradigm in black hole
physics.

As compared to some of the studies of dust collapse in higher 
dimensions mentioned above, we deal here 
within a general framework, i.e. there is no
restriction adopted on dimensions, the spacetime is not assumed to be 
self-similar, which is a somewhat restrictive condition for collapse
models, and
we treat marginally and non-marginally bound cases together by means
of a unified treatment. Also, we make no assumptions on the nature 
of the initial data, i.e. the initial density and velocity profiles
apart from their being regular, and these are taken to be 
suitably differentiable functions of the comoving radial coordinate
at the origin. Our results thus generalize earlier results on the topic
and provide a unified treatment. An additional advantage that
is derived due to the treatment here is that the 
dependence of the BH/NS end 
states on the nature of
initial data from which the collapse evolves, is brought out rather
clearly. The methodology used also allows us to consider 
in some detail the 
genericity and stability aspects related to the occurrence of naked
singularities in gravitational collapse.

To predict the final state of 
collapse for a given initial mass and velocity distribution, we study 
here the singularity curve
resulting from the collapse of successive matter shells, the tangent to 
which at the 
singularity is related
to the radially outgoing null geodesic equation. Hence, by determining the
sign of the tangent of the singularity curve at the central singularity we 
can integrate the null geodesics close to the singularity, and find 
whether a future directed radially outgoing null geodesic
comes out from the shell-focusing central singularity. 
This allows us to determine the final fate of collapse in terms 
of either a black hole, or a naked singularity.
It turns out that for
CCC to be true {\it even in higher dimensions}, one must impose a number of
constraints for the given system. The physical relevance of such 
assumptions is discussed here in some detail.

The paper is organized as follows. In the next section we set up the 
Einstein equations, and the regularity conditions for collapse are 
specified. In Section III, continually collapsing matter clouds are 
considered and it is shown how the initial data in terms of the initial
distributions of the density and velocity profiles determine the final
fate of the collapse in terms of either a black hole or a naked singularity. 
The dependence of these BH/NS phases on the nature of initial data is
brought out clearly here. In Section IV we discuss cosmic censorship
in higher dimensions from the perspective of our results, and the 
genericity and stability aspects related to naked singularity
formation are also considered in some detail. 
The case of marginally bound
collapse is discussed in some detail here together with some other
cases of interest, relating this to the question of
when it may be possible to restore the cosmic censorship for such models
in terms of certain possible physical assumptions for the initial density 
profiles. Some conclusions and remarks 
are summarized in the final Section V.

\section{Einstein Equations and regularity conditions}

To study the collapse of a spherical dust cloud, we choose 
a general spherically symmetric co-moving metric in $N\ge4$ dimensions 
which has the form,
\begin{equation}
ds^2=-dt^2+e^{2\s}dr^2+R^2(t,r)\dw
\label{eq:metric}
\end{equation}
where,
\begin{equation}
\dw=\sum_{i=1}^{N-2}\left[\prod_{j=1}^{i-1}\sin^2(\T^j)\right](d\T^i)^2
\end{equation}
is the metric on $(N-2)$ sphere. The energy-momentum tensor of the dust
has the form,
\begin{equation}
T^t_t=\rho(t,r);\;\;T^r_r=0;\;\;T^{\T^i}_{\T^i}=0
\end{equation}
We take the matter field to satisfy the {\it weak energy condition}, 
i.e. the energy density 
measured by any local observer be non-negative, and so for any 
timelike vector $V^i$, we must have,
\begin{equation}
T_{ik}V^iV^k\ge0\;\; \Longrightarrow\;\; \rho\ge0
\end{equation}
In the case of a collapsing cloud, the cloud has a finite boundary, $0<r<r_b$, 
outside which it is matched to a Schwarzschild exterior. The range of the 
coordinates for the metric is then $0<r<\infty$, and $-\infty<t<t_s(r)$
where $t_s(r)$ corresponds to the singular epoch $R=0$.

Using the above conditions, we can write the N-dimensional Einstein equations 
as 
~\cite{kn:rocha},
\begin{eqnarray}
(N-2)\left\{\frac{(N-3)}{2R^2}(1+\rdot^2-R^{'2}e^{-2\psi})+\right. 
&&\nonumber \\
\left.\frac{1}{R}(\dot{\psi}\rdot+e^{-2\psi}\psi'R'-e^{-2\psi}R'')
\right\}&=&\rho
\label{eq:ein1}
\end{eqnarray}
\begin{equation}
(N-2)\left\{\frac{N-3}{2R^2}(R{'2}e^{-2\psi}-\rdot^2-1)+\frac{\rddot}
{R}\right\}=0
\label{eq:ein2}
\end{equation} 
\begin{equation}
\frac{(N-2)}{R}\left\{2(\rdot'-\dot{\psi}R')\right\}=0
\label{eq:ein3}
\end{equation}
\begin{equation}
^NG^i_i\;\;(i=2,\cdots,N-2)=0
\label{eq:ein4}
\end{equation}
where $^NG^k_l$ is the N-dimensional Einstein tensor.
Integrating the first part of equation (\ref{eq:ein3}) we get,
\begin{equation} 
e^{2\psi}=\frac{R^{'2}}{1+\f}
\label{eq:tbl}
\end{equation}
where $\f$ is an arbitrary function of co-ordinate $r$, and $\f>-1$. 
Hence the 
generalized Tolman-Bondi-Lemaitre (TBL) metric in N-dimensions becomes,
\begin{equation} 
ds^2=dt^2-\frac{R^{'2}}{1+\f}dr^2-R^2(t,r)\dw
\label{eq:tbl1}
\end{equation}
Now substituting \e(\ref{eq:tbl}), in (\ref{eq:ein1}) and (\ref{eq:ein2}), 
we have,
\begin{equation}
\frac{(N-2)U'}{2R^{(N-2)}R'}=\rho,\;\;\;\frac{(N-2)\dot{U}}{2R^{(N-2)}\rdot}=0
\label{eq:tbl2}
\end{equation}
where we define and get by solving \e(\ref{eq:tbl2}),
\begin{equation}
U=R^{(N-3)}(\rdot^2-\f)\;\; ; \;\;\,U=\F 
\label{eq:u}
\end{equation}
Here $\F$ is another arbitrary function of the comoving
coordinate $r$. 
In spherically symmetric spacetimes 
$\F$ is the {\it mass function} which describes the mass distribution 
of the dust cloud 
and $\f$ is the {\it energy function} for the collapsing shells.
Thus using equations (\ref{eq:tbl2}) 
and (\ref{eq:u}), we finally get the required equations of motion as,
\begin{equation}
\frac{(N-2)F'}{2R^{(N-2)}R'}=\rho\;\;\; ; \;\;\; \rdot^2=\frac{\F}
{R^{(N-3)}}+\f
\label{eq:ein5}
\end{equation}
As seen from the above, once the mass function is
specified at the initial epoch
the energy function $\f$ fully specifies the velocity distribution
of infalling shells.
Also, the energy condition then implies $F'\ge0$.
It follows from the above that there is a spacetime singularity
at $R=0$ and at $R'=0$. The latter are called `shell-crossing 
singularities' which occur when successive shells of matter cross each 
other. These have not been considered generally to be genuine 
spacetime singularities, and possible extensions of spacetime 
have been investigated
through the same
~\cite{kn:clarke}.
On the other hand, the singularity at $R=0$ is where all matter 
shells collapse to a zero physical radius, and hence has been known
as a `shell-focusing singularity'. The nature of this singularity
has been investigated extensively in four-dimensional spacetimes
(see e.g. references in 
[4], 
and other reviews), and it is known for
the case of spherical dust collapse that it can be both naked or
covered depending on the nature of models one is considering.

The collapsing matter cloud condition implies that $\rdot<0$. 
For dust clouds it follows form the equations of motion that once 
$\rdot<0$ at the initial
epoch from where the collapse commences, then at all epochs we have
the same condition holding, and thus there is an endless collapse
till the shell-focusing singularity $R=0$ is reached.
In other words, there is no bounce possible in dust collapse
once the collapse has initiated with $\dot R <0$.
We use 
the scaling independence of the comoving coordinate $r$ to write
(see e.g. [4]),

\begin{equation}
R(t,r)=rv(t,r)
\label{eq:R}
\end{equation}
where,
\begin{equation}
v(t_i,r)=1\;\;\; ;\;\;\; v(t_s(r),r)=0\;\;\; ; \;\;\;\dot{v}<0
\label{eq:v}
\end{equation}
This means we have scaled the coordinate $r$ in such a way that 
at the initial epoch $R=r$ and at the singularity, $R=0$. 
It should be noted that we have $R=0$ both at the regular center $r=0$
of the cloud, and at the spacetime singularity where all matter shells
collapse to a zero physical radius. The regular center is then
distinguished from the singularity by a suitable behaviour of the 
mass function $F(r)$ so that the density remains finite and regular there
at all times till the singular epoch. 
The introduction of the parameter $v$ then allows us to distinguish 
the spacetime singularity from the regular center, with $v=1$ all through
the initial epoch, including the center $r=0$, which then decreases 
monotonically with time as collapse progresses to value $v=0$ 
at the singularity $R=0$.

From the equations of motion it is evident that to have a regular solution 
over all space at 
the initial epoch, the two free functions $\F$ and $\f$ must have the 
following forms,
\begin{eqnarray}
\F=r^{(N-1)}\M(r); & \f=r^2b(r)
\label{eq:forms}
\end{eqnarray}
where $\M(r)$ and $b(r)$ are at least $C^1$ functions 
of $r$ for $r=0$, and 
at least $C^2$ functions for $r>0$.
This is dictated by the condition that the density and energy
distributions must be regular at the initial epoch and should not be
blowing up. This is because if the mass function $F$ did not go as
power at least $r^{N-1}$ closer to the origin, then as implied from
the equations of motion, the density will
be singular at the origin $r=0$ as it will diverge there, and that 
cannot be accepted
as regular initial data for collapse. Similarly, equation (12) implies 
that $f(r)$ is determined once the velocity profile is specified, and
vice-versa, for a given initial density distribution. Since the center
of the cloud is taken to be at rest in any spherically symmetric 
distribution, the leading term in the energy profile must be at least
$r$ or higher. Then again equation (12) implies the behaviour for
$f(r)$ as above.

\section{Continual Collapse and BH/NS End States}

We now consider the endless collapse of a dust cloud 
to a final shell-focusing singularity at $R=0$, where
all matter shells collapse to a zero physical radius. In particular, 
we analyze specifically the nature of the central singularity at $R=0,r=0$   
in detail to determine when it will be covered by the event horizon,
and when visible and causally connected to outside observers. If there
are future directed families of non-spacelike curves, coming out from 
the singularity and reaching faraway observers, then the singularity
will be naked. The absence of such families will give a covered
case when the result is a black hole.

In the following, a higher dimensional collapse as discussed 
above is considered
and we shall show how the initial density and energy distribution 
prescribed at the initial epoch from which the collapse commences, 
completely determine if there will be families of non-spacelike 
trajectories coming out of the singularity. We shall show that for
a generic situation, given an initial density distribution for the
collapse to develop, it is always possible to choose an energy
profile so that the collapse of this density profile ends in a naked
singularity, or one could also choose another class of energy distribution
so that the same density distribution will end up collapsing so as
to create a black hole. That is, given an initial density profile,
the outcome in terms of either a black hole or a naked singularity
as end states really depends on the class of the energy distribution
chosen. The 
converse will also be seen to hold true,
that is, given the initial energy function, one will be able to
choose classes of density profiles, subject to the the weak energy condition, 
so as to give rise to either of the BH/NS end states depending on
the choice made. Our results reduce to
the usual dust collapse models of general relativity when 
$N=4$.

With the regular initial conditions as above, \e(\ref{eq:ein5}) 
can be written as,
\begin{equation}
v^{\frac{N-3}{2}}\dot{v}=-\sqrt{\M(r)+v^{(N-3)}b(r)}
\label{eq:dotv}
\end{equation}
Here the negative sign implies that $\dot{v}<0$, that is, the matter 
cloud is collapsing. Integrating the above equation with respect to $v$, 
we get,
\begin{equation}
t(v,r)=\int_v^1\frac{v^{\frac{N-3}{2}}dv}{\sqrt{\M(r)+v^{(N-3)}b(r)}}
\label{eq:scurve1}
\end{equation}
We note that the co-ordinate $r$ is to be treated as a constant in the above 
equation. Expanding $t(v,r)$ around the center, we get,
\begin{equation} 
t(v,r)=t(v,0)+r\X(v)+r^2\frac{\X_2(v)}{2}+r^3\frac{\X_3(v)}{6}+\cdots
\label{eq:scurve2}
\end{equation}
where the function $\X(v)$ is given by,
\begin{equation}
\X(v)=-\frac{1}{2}\int_v^1\frac{v^{\frac{N-3}{2}}(\M_1+v^{(N-3)}b_1)dv}
{(\M_0+v^{(N-3)}b_{0})^{\frac{3}{2}}}
\label{eq:tangent}
\end{equation}
where,
\begin{equation}
b_{0}=b(0)\;\; ;\;\;\M_0=\M(0)\;\; ;\;\;b_1=b'(0)\;\; ;\;\;\M_1=\M'(0)
\label{eq:notation}
\end{equation}
Thus, the time taken for the central shell to reach the singularity is 
given as
\begin{equation}
t_{s_0}=\int_0^1\frac{v^{\frac{N-3}{2}}dv}{\sqrt{\M_0+v^{(N-3)}b_{0}}}
\label{eq:scurve3}
\end{equation}
From the above equation it is clear that for $t_{s_0}$ to be defined,
\begin{equation}
\M_0+v^{(N-3)}b_{0}>0
\label{eq:constraint}
\end{equation}
In other words, the continual collapse condition
implies the positivity of the above term.
Hence the time taken for other shells to reach the singularity 
can be given by the expansion,
\begin{equation}
t_s(r)=t_{s_0}+r\X(0)+{\cal O}(r^2)
\label{eq:scurve4}
\end{equation}
Also, from \e(\ref{eq:dotv}) and (\ref{eq:scurve2}) we get for small 
values of $r$, along constant $v$ surfaces,
\begin{equation}
v^{\frac{N-3}{2}}v'=\sqrt{(\M_0+v^{(N-3)}b_{0})}\left(\X(v)+r\X_2(v)+
\cdots\right)
\label{eq:vdash}
\end{equation}
Now we can easily see that the value of $\X(0)$ depends on and is completely
characterized by the functions 
$M(r)$ and $b(r)$, which in turn specify fully the initial mass and 
energy
distributions for the collapsing matter. Specifying these functions 
is equivalent to specifying the regular initial data for collapse on the 
initial surface $t=0$. In other words, a given set 
of density and energy distribution completely determines the slope to the 
singularity curve at the origin, which is the central singularity. 
Also, it is evident that given any one of these two profiles we can 
always choose the other one in such a manner so that 
the quantity $\X(0)$ will be either positive or negative.

In order to determine the visibility or otherwise of the central
singularity, we now need to analyze the behaviour of non-spacelike 
curves in the vicinity of the singularity and the causal structure 
of the trapped surfaces.

The boundary of the trapped surface region of the space-time
is given by the apparent horizon within the collapsing cloud, which
is given by the \e,
\begin{equation}
\frac{F}{R^{N-3}}=1
\label{eq:apphorizon} 
\end{equation}
What we need to determine now is when there will be families
of non-spacelike paths coming out of the singularity, reaching 
far away observers, and when there will be none. 
The visibility or other wise of the singularity is decided
accordingly.
Broadly, it can be stated that if the neighborhood of the center 
gets trapped earlier than the singularity, then it is covered, 
otherwise it is naked with 
families of non-spacelike future directed trajectories escaping away from it.
By determining the nature of the singularity
curve and its relation to the initial data, we are able to deduce
whether the trapped surface formation in collapse takes place
before or after the singularity. It is this causal
structure that determines the possible emergence or otherwise of
non-spacelike paths from the singularity, and settles 
the final outcome in terms of either a BH or NS.

To consider the possibility of existence of such families, and 
to examine the nature of the singularity occurring at $R=0$, $r=0$ for the 
scenario under consideration, consider the outgoing 
radial null geodesics equation,
\begin{equation}
\frac{dt}{dr}=e^{\psi}
\label{eq:null1}
\end{equation}
The singularity occurs at a point $v(t_s(r),r)=0$, which corresponds 
to $R(t_s(r),r)=0$. Therefore, if we have any future directed null geodesics 
terminating in the past at the 
singularity, we must have $R\rightarrow0$ as $t\rightarrow t_s$. 
Now writing \e (\ref{eq:null1}) in terms of variables 
$(u=r^\alpha,R)$ where $\alpha>1$, we have,
\begin{equation}
\frac{dR}{du}=\frac{1}{\alpha}r^{-(\alpha-1)}R'\left[1+\frac{\dot{R}}
{R'}e^{\psi}\right]
\label{eq:null2}
\end{equation}
Choosing $\alpha=\frac{N+1}{N-1}$, and using \e (\ref{eq:ein5}) 
together with the collapse condition $\dot{R}<0$, we get,
\begin{equation}
\frac{dR}{du}=\frac{N-1}{N+1}\left(\frac{R}{u}+\frac{v'v^{\frac{N-3}{2}}}
{(\frac{R}{u})^{\frac{N-3}{2}}}\right)\left(\frac{1-\frac{F}{R^{N-3}}}
{e^{-\psi}R'\left(e^{-\psi}R'+\left|\rdot\right|\right)}\right)
\label{eq:null3}
\end{equation}
If null geodesics terminate at the singularity in the past with a 
definite tangent, 
then at the singularity we have $\frac{dR}{du}>0$, 
in the $(u,R)$ plane which must have a finite value.

In the case under consideration, all 
singularities for $r>0$ are covered since $\frac{F}{R}\rightarrow\infty$ 
in the limit of approach to the singularity in that case, and hence 
$\frac{dR}{du}\rightarrow-\infty$. Therefore only the singularity at 
the central shell could be naked.

In order to see possible emergence of null geodesics
from the central singularity, we now need to analyze  \e (\ref{eq:null3}).
Let us calculate the limits of the concerned functions in \e 
(\ref{eq:null3}) at the central singularity. From \e (\ref{eq:tbl}) 
we get that in the limit of $t\rightarrow t_s,r\rightarrow 0$ we have
$e^{-\psi}R'\rightarrow 1$. Also from \e (\ref{eq:dotv}) and
(\ref{eq:vdash}), we have in this limit $\dot{R}\rightarrow 0$.
It then follows in general, from the Einstein equations
discussed above, that the term $F/R^{N-3}$ goes to 
zero in this limit.

We would like to find when there will be future directed null 
geodesics coming out from the central singularity with a well-defined 
and definite positive tangent
in the $(t,r)$ or $(R,u)$ plane, thus making the singularity
visible. 
Let us define the tangent to the null geodesic at the singularity as,
\begin{equation}
x_0=\lim_{t\rightarrow t_s}\lim_{r\rightarrow 0} \frac{R}{u}=\left.\frac{dR}{du}
\right|_{t\rightarrow t_s;r\rightarrow 0}
\end{equation}
Using equations (\ref{eq:null3}) and (\ref{eq:vdash}), 
along with the
required limits as above, we get,
\begin{equation}
x_0^{\frac{N-1}{2}}=\frac{N-1}{2}\sqrt{\M_0}\X(0)
\label{eq:x01}
\end{equation}

Let us now deduce the necessary and sufficient conditions for
a naked singularity to exist, that is, for null geodesics with a 
well-defined tangent to come out from the central singularity. 
Suppose we have $\X(0)>0$, then we 
always have (from \e (\ref{eq:x01})), $x_0>0$ and 
then in the $(R,u)$ plane, the equation for the null geodesic
that comes out from the singularity is given by 
\begin{equation}
R=x_0u
\label{eq:x02}
\end{equation}
In other words, \e (\ref{eq:x02})
is a solution of the null geodesic equation in the limit of
the central singularity. Thus given $\X(0)>0$, we can always construct
a solution of radially outgoing null
geodesics emerging from the
singularity. This makes the central singularity visible.
In the $(t,r)$ plane, the null geodesics above near 
the singularity will be given as,
\begin{equation}
t-t_s(0)=x_0r^{\frac{N+1}{N-1}}
\end{equation}
It follows that $\X(0)>0$ implies $x_0>0$ and we get radially 
outgoing null 
geodesics emerging from the 
singularity, giving rise to the central naked singularity.

On the other hand, if $\X(0)<0$, then we see that the singularity
curve is a decreasing function of $r$. Hence the region around the center
gets singular before the central shell, and after that it is no more 
in the spacetime. Now if there would have been any {\it outgoing} 
null geodesic from the central singularity, it must then go to 
a singular region or outside the spacetime, which is impossible. 
Hence when $\X(0)<0$, we always have a black hole solution.

If $\X(0)=0$ then we will have to take into account the next higher 
order non-zero term in the singularity curve equation, and 
do a similar analysis by 
choosing a different value of $\alpha$ in 
equation (\ref{eq:null2}).

We have thus shown above that $\X(0)>0$ is the necessary and
sufficient condition for null geodesics to come out from the central
singularity with a definite positive tangent.
It should be noted, however, that in general the dependence of $R$
on $r$ along the outgoing null
geodesics from the singularity does not necessarily have to be of
a power-law form.
However, in order to satisfy the regularity and physical relevance,
examining the trajectories which come out with a regular and well-defined
tangent is physically more appealing, which is the case we have
examined here.

We show below for completeness, however, that if null geodesics 
of any form come out at all, then those with definite tangent also must 
emerge from the central singularity.
Towards this, let us consider the equation of apparent horizon, which is
from equations (\ref{eq:apphorizon}) and (\ref{eq:scurve1})
given by,
\begin{equation}
t_{ah}(r)=t_{s_0}+r\X(0)+r^2\frac{\X_2(0)}{2}+\cdots - 
{\cal O}(r^{\frac{N-1}{N-3}})
\label{eq:apphorizon2}
\end{equation}
Since the apparent horizon is a well-behaved surface 
as one initiates close to $r=0$, for a spherical
dust collapse, hence we can say that the singularity curve
for the collapse and the derivatives around the center are also
well-defined, as the same coefficients are present in both 
(\ref{eq:apphorizon2}) and (\ref{eq:scurve2}).
Also, this shows that whenever $\X(0)$ is negative, the region
around the center gets trapped before the central singularity,
giving a sufficient condition for a black hole to develop. 
It follows that if null geodesics are coming out, then
at least one of the coefficients $\X$ must be non-vanishing
and positive. Then, as we have already shown, null geodesics
with definite tangent will come out from the central singularity.

From the above it follows that {\it in the absence 
of a null geodesic with a definite tangent there cannot be any 
null geodesics coming out of the singularity.}

It is also clear now, from equation (20), that
whether $\X(0)>0$ or otherwise is fully determined by the regular 
initial data for collapse, in terms of the given initial density
and energy distribution for the collapsing shells. It thus follows
that the initial data here completely determines the final fate of 
collapse in terms of BH/NS end states. We shall discuss this
further in the next section.

\section{Cosmic censorship and genericity issues}

It is now possible to examine the question of validity  
of cosmic censorship in a higher dimensional collapse scenario under 
consideration. Since we can explicitly find out as pointed out above,
when families of non-spacelike geodesics can come out of the singularity
or otherwise, we can address now this question below in some detail, 
and we also discuss some related issues.

To focus the discussion, let us consider a model initial density profile 
as given by,
\begin{equation}
\rho(t_i,r)=\rho_0+r\rho_1+r^2\frac{\rho_2}{2!}+r^3\frac{\rho_3}{3!}+\cdots
\label{eq:denprofile}
\end{equation}  
and we write the function $\M(r)$ as,
\begin{equation}
\M(r)=\sum_{n=0}^{\infty}\M_nr^n\;\;\; ;\;\;\;\M_n=
\frac{2\rho_n}{(N-2)(N+n-1)n!}
\label{eq:mn}
\end{equation}
along with an energy profile as specified by,
\begin{equation}
b(r)=b_0+rb_1+r^2\frac{b_2}{2!}+\cdots
\end{equation}

\subsection{Marginally bound collapse}

In the first place, let us consider the class of {\it marginally bound} 
collapse models for a more 
transparent understanding of the problem.
This is the case when the energy function $b(r)$ above vanishes 
identically for the collapsing shells. In such a situation, the
first non-vanishing 
coefficient $\X_n(0)$,where $n>0$, as described in equation 
(\ref{eq:scurve2}) are given by,
\begin{equation}
\X_n(0)=-\frac{n!}{N-1}\left(\frac{\M_n}{\M_0^{\frac{3}{2}}}\right)
\label{eq:xn}
\end{equation}
The quantities $\M_n$ are described in equation (\ref{eq:mn}). 
Now it is evident that whenever $\rho_1<0$, we will get a naked singularity 
{\it in all dimensions}, whereas $\rho_1>0$ always results in a black 
hole. The case $\rho_1<0$ corresponds to the physical situation when
the density decreases with increasing comoving radius $r$, as one 
would typically expect the density to be highest at the center and
then gradually decrease as we move out in any realistic configuration
such as a massive star. Further, we note that the above conclusion is 
not dependent on the magnitude of  $\rho_1$, but only on its sign,
that is density must decrease away from the center with the density
gradient being non-zero. Thus it becomes clear that it is the 
{\it density inhomogeneity} that delays the formation the trapped surfaces,
thus causing a naked singularity. This is closely connected to the 
non-vanishing {\it spacetime shear}, and for a discussion of how shear
distorts the geometry of trapped surfaces closer to the spacetime 
singularity, we refer to 
~\cite{kn:roy}.

Let us, however, assume that the initial density 
distribution has all odd terms in $r$ vanishing, i.e. it admits 
no `cusps' at the center and that it is either sufficiently 
differentiable, or is a smooth and analytic function of $r$. In that 
case, we must have $\rho_1=0$. Then we get from equation (\ref{eq:vdash}) 
that in the neighborhood of the singularity, the behavior of $v$ is
given by,
\begin{equation}
\lim_{t\rightarrow t_s}\lim_{r\rightarrow 0} v=\left[\frac{N-1}{4}
\sqrt{\M_0}\X_2(0)\right]^{\frac{2}{N-1}}r^{\frac{4}{N-1}}
\label{eq:vr}
\end{equation}
Also, in the same limit the function $\frac{F}{R^{N-3}}$ has the form
\begin{equation}
\lim_{t\rightarrow t_s}\lim_{r\rightarrow 0}\frac{F}{R^{N-3}}=
\frac{r^2\M_0}{v^{(N-3)}}
\label{eq:applimit}
\end{equation}
Thus it is clear from equations (\ref{eq:vr}) and (\ref{eq:applimit}), 
that if $N>5$ then for $\lim_{t\rightarrow t_s},\lim_{r\rightarrow 0}$, 
$\frac{F}{R}\rightarrow\infty$ and thus the end state of collapse
will always be a black hole (see also 
[10]). 
It thus follows that for a marginally bound dust collapse, with 
$\rho_1=0$, i.e. when the initial density profile is sufficiently 
differentiable and smooth, the CCC is always respected in a higher dimensional 
spacetime with $N=6$ or higher.

\subsection{Critical value of $\rho_2$ in five-dimensional case}

We note that in the conclusion above, the spacetime dimension has to be
either six, or higher. Let us consider the case when the spacetime dimension
is five, but still with an analytic initial density profile.

In case of a five-dimensional marginally bound collapse with $\rho_1=0$, 
we can write the tangent to the outgoing radial null geodesic at the
singularity in the $(R,u)$ plane as,
\begin{equation}
x_0^2=\sqrt{\M_0}\X_2(0)\frac{\left[1-\sqrt{\frac{F}{R^2}}\right]}
{\left[1+\sqrt{\frac{F}{R^2}}\right]}
\end{equation}
The sufficient condition for the existence of an outgoing null geodesic 
from the singularity is
we must have $x_0>0$, which in the above case amounts to,
\begin{equation}
\xi \equiv \frac{\M_2}{\M_0^2}<-2
\label{eq:condition1}
\end{equation} 
But again, the outgoing null geodesic should be within the spacetime 
{\it i.e} the slope of the geodesic must be less than that of the 
singularity
curve,
\begin{equation}
\lim_{t\rightarrow t_s}\lim_{r\rightarrow 0}\left(\frac{dt}{dr}\right)_{null}
\le \left(\frac{dt}{dr}\right)_{sing}
\end{equation}
From the above equation we get the condition,
\begin{equation}
\xi=\frac{\M_2}{\M_0^2}\le-8
\label{eq:condition2}
\end{equation}
Thus from equation (\ref{eq:condition1}) and (\ref{eq:condition2}) we see 
that for an outgoing null geodesic from the singularity to exist we must 
have $\xi\le\xi_c=-8$,
in which case we get a naked singularity, and otherwise a black hole
results as collapse end state.

We note that this situation has an interesting parallel to the 
four-dimensional collapse
scenario, where we have a similar critical value existing, however,
it is for the coefficient $\rho_3$, when both $\rho_1$ and $\rho_2$ 
are vanishing 
~\cite{kn:initial}. 
Thus, with the increase of the spacetime
dimension by one, the criticality separating the BH/NS phases shifts at 
the level of second density derivative from the earlier third
density derivative.

An interesting observation we could make here is that
for $\xi<-2$ we have an 
increasing apparent horizon at the singularity.
The apparent horizon is given by $R=F$ so it initiates
at the central singularity $r=R=0$, and in the
above case it is increasing in time (as opposed to
the Oppenheimer-Snyder case of homogeneous dust collapse).
Thus for the range $-2>\xi>-8$ no trapped surface is formed till the 
singularity epoch, but still we get a black hole as the collapse 
end state. This confirms that {\it the absence of a trapped surface 
till the singularity is necessary, but not a sufficient condition 
for the formation of a naked singularity}. 

\subsection{Marginally bound collapse with initial homogeneous density}

It is useful to note here that the well-known Oppenheimer-Snyder 
class of collapse solutions is a special case of marginally 
bound dust collapse in four dimensions in which 
the initial density profile is homogeneous, that is, $\M_n (n>0)=0$
for all $n$. The point is, if the initial density is homogeneous
but if the collapse is {\it not} marginally bound, then the
non-zero energy function $f$ could inhomogenize the collapse
at later epochs. In the present case, as $f=0$, at all 
later epochs also the density remains a function
of time only, that is, it is homogeneous at all later times
as well.
Thus we clearly see that the final outcome of this class of collapse is 
always a black hole.
Furthermore from \e s (\ref{eq:scurve4}) and (\ref{eq:scurve2}) we see that,
\begin{equation} 
t_s(r)=t_{s_0}\;\;\; ;\;\;\; v(t,r)=v(t)
\end{equation}
As the scale function $v$ is independent of $r$, all the shells
collapse simultaneously to the singularity. The time taken to 
reach the singularity is given as,
\begin{equation}
t_{s_0}=\frac{2}{(N-1)\M_0^{\frac{1}{2}}}
\end{equation}
Thus, as we go to higher dimensions, for a given density, the 
time taken to reach the singularity will reduce.

It is, however, interesting to note that even if we start
with an initial homogeneous density profile but allow for
non-zero initial radial and tangential pressures of the form,
\begin{equation}
p_{r}(t,r)=1\;\; ;p_{\theta_0}(r)=1+p_{\theta_2}r^2+p_{\theta_3}r^3
+\cdots \end{equation}
then as shown in ~\cite{kn:gospsj}, we have
\begin{equation}
\X(0)=-\frac{1}{3}\int_0^1\frac{v^{\frac{N+5}{2}}(p_{\theta_3})}
{v^{(N-1)}(p_{\theta_2}-\frac{1}{3})+\frac{2}{3}}
\end{equation}
Thus it is seen that a negative $p_{\theta_3}$ coefficient
does lead to a naked singularity.

This is similar to the case of a collapse 
which is not marginally bound, where an initially homogeneous 
density profile can turn inhomogeneous at later epochs due to 
the non-vanishing shell velocities. In the same way, in the case 
above also the non-vanishing pressures could inhomogenize 
the initially homogeneous density distribution at later epochs 
to cause a naked singularity 
eventually. It has to be noted, all the same, that the equation of 
state in the situations such as above could be considered to be
some what peculiar (though matter is fully normal, satisfying the
positivity of energy condition, and collapse conditions are fully
regular). To state this differently, one can argue that the models
where only purely tangential pressures are taken to be non-vanishing
may not be considered to be physically realistic, and if we choose 
the equation
of state to be say $p=k\rho, k>0$, or any homogeneous equation of
state, then when the initial density profile is taken to be homogeneous,
then so will be the initial pressures, and then the collapse
will end up in a black hole only, and no naked singularity 
will arise.

\subsection{Genericity and stability of naked singularities}

In the above, we have discussed various special subcases of
a higher dimensional collapse scenario which result either in a 
black hole or a naked singularity end state, depending on values and 
behaviour of the parameters involved, and we also discussed the
possible preservation of CCC under various set of assumptions.

It is necessary however, to look at the situation in a 
collective manner if we are to gain any insight on the genericity 
and stability aspects connected to the naked singularities forming in 
gravitational
collapse. It is well-known of course, that the genericity and stability
are quite involved issues in the general theory of relativity, and that
there
does not exist any well-defined way to test the same in a unique manner
for a given situation. Also, there may be different kinds of stabilities
involved. For example, we can ask here, if the conclusions above will
be stable to non-spherical perturbations, or when forms of matter more
general than dust are considered and so on. Such issues are worth a 
separate and
detailed investigation, and will be crucial towards the important
problem of collapse end states.

At a somewhat different, but still quite interesting level, we can 
inquire about the stability of BH/NS end states with respect to the
perturbations in the initial data space which determines the 
final outcome of collapse
as we have seen above. As pointed out here, this is a function
space consisting of all possible mass functions $F$ and energy functions
$f$. It is worth knowing how, for example, a naked singularity end state
would be affected when one moves from a given density and energy profile
(which gave rise to this state) to a nearby density or energy profile
in this space of all initial data.

The issue of how given density and energy distributions determine
the final collapse end state has been discussed quite extensively in the
usual four-dimensional dust collapse models, though using a somewhat
different methodology 
~\cite{kn:initial}. 
These results were completed
to give a full and general treatment of four-dimensional case in
~\cite{kn:Jhijo}, 
and the typical result is that given any density 
profile, one could choose the energy profile (and vice-versa), so that
the collapse end state would be either a black hole, or a naked 
singularity depending on this choice.

Our results here generalize this to the case of a higher 
dimensional collapse situation, and our method now allows us 
to make a more definite statement on genericity of naked singularity 
formation (in the sense stated above). As we see from equation (20), 
the quantity $\X(0)$ is fully determined from the initial data functions 
and their first derivatives. Once it is positive, the collapse ends in a 
naked singularity and a negative value gives black hole. It follows by 
continuity that given a density profile, if the
energy profile chosen is such that the collapse ends in NS, 
i.e. $\X(0)>0$, then there is a whole family of near by velocities 
such that this will continue to be the case, and the NS will 
form an open subspace in the 
initial data space. Same of course holds for BH formations 
as well, and both these are neatly separated open regions 
in the initial data space. But if we take on physical grounds 
both $\rho_1$ and $b_1$ to be 
vanishing, then from equation (20) we have $\X(0)=0$,
and CCC may be restored.       
 

\section{Concluding remarks}

In this final section, we give several concluding remarks
and observations.

1. It follows that if we can suitably motivate physically all 
the assumptions such as those discussed above, then it may be possible 
to restore CCC in a higher dimensional spacetime. Let us discuss 
such conditions in some detail. That the equation of state must be 
dust-like in the final phases of collapse is a strong
assumption, but it is not a possibility that can be completely 
ruled out (see e.g. 
[13]).
After all, we know very little 
on the equations of state, especially how it would be like in the 
advanced stages of collapse. Also, it is quite possible that in 
the very late stages of collapse the configuration
is very much like a marginally bound one in the vicinity of 
the singularity. The introduction of pressures may or may not change 
such a scenario.

All the same, in our view, the assumption $\rho_1=0$ is a 
tricky one, and this has been extensively
discussed in the past within the context of collapse in
four-dimensions. While it may be 
quite convenient to deal 
with smooth and analytic density profiles, especially when it 
comes to numerical 
models, it should not be forgotten that after all this is only an extra
assumption, and that the basic equations of general relativity do not
demand any such constraint. Neither it is clear astrophysically that the 
interiors of
the stars must necessarily have analytic density or energy 
distribution. In certain
equilibrium cases, the field equations imply that these have to be smooth,
but this need not be true in general, and especially the dynamically 
developing 
collapse situations could be quite different.

We thus conclude that each of the above
assumptions require further scrutiny and sufficient physical motivation
so as to arrive at any definite conclusion on the status of CCC in a 
higher dimensional spacetime. However, this will be certainly worth
the effort, given the importance of these issues in
black hole physics.

2. Let us consider the scenarios when some of 
the above assumptions break down. We have already noted earlier 
that when $\rho_1$ is non-vanishing, then the collapse ends in 
a naked singularity in all dimensions, including $N=4$. Again, 
our considerations above immediately imply that whenever spacetime 
is {\it not} marginally bound, the collapse always results into both 
the BH/NS phases as collapse end states, depending on the nature of 
initial data, {\it irrespective of $\rho_1$ being either zero or 
non-zero}. That is, in a 
more generic non-marginally bound case, the condition 
$\rho_1=0$ does not save the CCC. However, if we take on physical
grounds both $b_1=\rho_1=0$ then from equation (20) $\X(0)=0$
and CCC could be preserved.

3. Consider the situation when we must believe some how that 
in the later stages of collapse the form of matter cannot be dust-like, 
and that non-dust forms of matter, and effects of pressures must be 
suitably taken into account. In such a case, as pointed out above,
it is seen that even 
if one considered only homogeneous initial 
density profiles (however, with non-zero initial pressures), then also
the pressure by it-self can cause sufficient distortions in the 
formation of the apparent horizon so as to cause a naked 
singularity as end state of collapse, rather than a black hole. 
However, if the equation of state is homogeneous, together with
initial data being homogeneous density profile, then no
naked singularity will appear.

4. We would like to suggest that there may be some hope, as 
outlined above, 
to recover CCC while we transit to a higher dimensional spacetime arena.
This is subject to validity of several extra physical inputs as we 
described above.                              
On the other hand, once we move to more general situations of either a 
non-marginally bound case, or with a more general form of matter, or without
restrictive extra-assumptions on the nature of the initial density 
profiles, 
then generically both the BH/NS phases could result as end states 
of collapse in a higher dimensional spacetime scenario.    
It would be fair to state that dynamical collapse in general 
relativity offers a rich spectrum of possibilities to
investigate.

5. Finally, we point out that the formalism given here brings
out the role of initial data in causing BH/NS end states  
for the four-dimensional dust collapse also in a clear and transparent
manner, completing earlier results in this direction. To be specific,
it is seen, using equation (20), that given any initial density    
distribution for the cloud, one can always choose a suitable energy
profile so that the evolution could end in either of a black hole
or a naked singularity depending on the choice made. In other words,
there is a non-zero measure of energy distributions which will take
the given density profile to a black hole, and the same holds for
a naked singularity to evolve from the same initial density.
The converse is also true, namely, given any initial energy
distribution, one can choose the density profiles which give
rise to either of these end states.

Acknowledgment: It is a pleasure to thank the referee for
helpful comments.

\end{document}